\documentclass[twocolumn]{aastex631}
\usepackage{xcolor}
\usepackage{soul}
\usepackage[normalem]{ulem}
\usepackage{amsmath}
\newcommand{\wvap}[0]{$X(\mathrm{H_2O})$}
\newcommand{\nwat}[0]{$N(\mathrm{H_2O})$}
\newcommand{\nice}[0]{$N(\mathrm{sH_2O})$}

\accepted{June 5, 2024}
\submitjournal{ApJ}

\begin{document}

\title{Cold water emission cannot be used to infer depletion of bulk elemental oxygen [O/H] in disks}

\author{Maxime Ruaud}
\affiliation{NASA Ames Research Center, Moffett Field, CA, USA}
\affiliation{Carl Sagan Center, SETI Institute, Mountain View, CA, USA}

\author{Uma Gorti}
\affiliation{NASA Ames Research Center, Moffett Field, CA, USA}
\affiliation{Carl Sagan Center, SETI Institute, Mountain View, CA, USA}

\begin{abstract}
We re-examine the constraints provided by \textit{Herschel} Space Observatory data regarding 
cold water emission from protoplanetary disks. Previous disk models that were used to interpret observed water emission concluded that  oxygen (O/H) is depleted by at least 2 orders of magnitude if a standard, interstellar gas/dust mass ratio is assumed in the disk. In this work, we use model results from a recent disk parameter survey and show that most of the \textit{Herschel} constraints obtained for cold water (i.e. for transitions with an upper energy level $E_\mathrm{up}<200$ K, where the bulk of the disk water lies) can be explained with disk models adopting ISM-like oxygen elemental abundance (i.e. O/H=$3.2\times10^{-4}$) and the canonical gas/dust mass ratio of 100. We show that cold water vapor is mainly formed by photodesorption of water ice at the interface between the molecular layer and the midplane, and that its emission is relatively independent of the main disk properties like the disk gas mass and gas/dust mass ratio. We find that the abundance of water vapor in the outer disk is set by photoprocesses and depends on the (constant) vertical column density of water ice needed to attenuate the FUV photon flux, resulting in roughly constant emission for the parameters (gas mass, dust mass, disk radius) varied in our survey. Importantly, water line emission is found to be optically thick and hence sensitive to temperature more than abundance, possibly driving previous inferences of large scale oxygen depletion. 
\end{abstract}

\section{Introduction} 
\label{sec:intro}
The dominant mass reservoir of water in disks is likely in the form of water ice \citep[e.g.][]{Pontoppidan14,Oberg21}. Water in the vapor form  is predicted to be abundant in three main regions of disks \citep[see e.g.,][]{Woitke09,vanDishoeck21}. A first reservoir is located inside the water snowline near the midplane where water abundance is high (i.e. $\sim 10^{-4}$) but not accessible to observations because of the high dust opacity at mid-IR wavelengths. A second reservoir is located in the warm surface layers of the inner disk where high temperature chemistry dominates its formation and has been observed at mid-IR wavelengths using Spitzer \citep{Pontoppidan10}, and for which JWST is currently providing additional constraints \citep{Tabone23,Banzatti23}. Finally, a third, and the most massive reservoir is located in the outer disk ($r \gtrsim$ a few au)  and close to the midplane where water is efficiently photodesorbed from the ice phase by stellar and interstellar UV photons \citep[e.g.][]{Du14,Ruaud19}. Due to the low disk temperatures, most of this emission is expected to arise at far-infrared and sub-millimeter wavelength regions. 

Important, although limited, constraints have been obtained on this cold water reservoir from Herschel-HIFI through the WISH survey \citep{vanDishoeck21}, the Herschel-PACS GASPS \citep{Alonso17} and DIGIT \citep{Fedele13} surveys, and data from other observational programs described in \citet{Pontoppidan10,Blevins16}. Water emission was found to be far lower than anticipated. In particular, observations of the two ground state rotational transitions; o-H$_2$O $1_{1_0}-1_{0_1}$ at 538.29$\mu m$ and p-H$_2$O $1_{1_1}-1_{0_0}$ at 269.27$\mu m$, in 14 disks (i.e. 10 disks around T Tauri stars and 4 disks around Herbig Ae stars) revealed much weaker emission than predicted by theoretical models \citep{Hogerheijde11,Du17}. In fact, out of the fourteen observed disks, only three yielded detection (TW Hya, HD 100546 and DG Tau) despite deep integration times of $\gtrsim 10$ hours. For all the other sources, only upper limits were obtained \citep[see][]{Du17}. 

These observations led \citet{Du17} to suggest that the gas phase oxygen abundance in these disks is decreased by at least 2 orders of magnitude  compared to the typical oxygen abundance found in the interstellar medium (ISM) if the same value of the dust-to-gas mass ratio ($\epsilon = 10^{-2}$) is employed. In order to bring about this reduction, they propose that the amount of oxygen that is available to the chemistry of the disk decreases over time due to the settling of icy grains in the midplane \citep{Bergin10,Hogerheijde11} and/or due to their incorporation into larger bodies that do not contribute to disk chemistry \citep[e.g.][]{Krijt16,Du17}. 

A similar scenario---incorporation of ices into larger planetesimal sinks---has been invoked for CO, the other main O reservoir. In fact, large depletions of both C and O have been claimed to explain the discrepancies between observed fluxes of CO and its isotopologues in disks compared to model predictions\citep{Du15,Kama16,Schwarz16,Krijt16,Krijt18,Krijt20,Zhang19,Calahan21,Schwarz21}. However, in recent work, we call this scenario into question \citep{Ruaud22} and  show that more sophisticated models can explain observed CO isotopologue line emission from surveys to within a factor of $\sim 2-3$ without invoking strong depletion factors of C and O and moreover, using a normal interstellar dust/gas ratio ($\epsilon = 10^{-2}$). In related work, atomic carbon emission ([CI] 492GHz) observed with ALMA was also found to be consistent with our models \citep{Pascucci2023}. The lack of depletion of C and O has implications for other lines like hydrocarbons (C$_2$H and C$_3$H$_2$) and cyanides (CN and HCN) which often show brighter emission than expected. The aforementioned sequestration of O into ice-laden planetesimals could still be a viable explanation, but large scale depletions are not necessary and a factor of $\sim$ 2 change in the C/O ratio may be all that is needed to reconcile the carbon chemistry \citep[e.g.][]{Bergin16,Ruaud22}. 

However, one of the issues with tracers like hydrocarbons and cyanides is that even though they are bright in disks, they are not reservoir-species; i.e. their abundance in disks are orders of magnitude lower than their constituent elemental abundances. Therefore, modeled emissions are much more sensitive to the assumed chemical pathways and rates, which can be uncertain. Furthermore, important chemical processes that affect the predicted abundances of these trace species are often neglected or approximated in disk models, for example, chemistry involving excited H$_2$ which has a considerable impact on the chemistry of hydrocarbons and cyanides at the disk surface \citet{Cazzoletti18,Ruaud21}. Finally, it has long been suggested that local gas-phase carbon enhancements (rather than O depletion) could also cause an increase in the C/O ratio and affect carbon chemistry \citep[e.g.,][]{kastner2015, anderson2017, wei2019}. 

Water, on the other hand, is less affected by such complications and water chemistry is relatively well understood. Most of the oxygen in disks is believed to be in the form of water ice in the midplane, and observed gas phase water most likely originates from desorption of this ice. Even though emission of cold water in disks traces only a very small fraction of the total available oxygen, the vapor potentially tracks the main water ice reservoir, and its emission may be used to investigate changes in the O/H ratio in disks. In this paper, we use existing observational constraints obtained from cold water lines with Herschel (i.e. transitions with an upper energy level $E_\mathrm{up} < 200$ K) to further investigate our earlier results obtained for CO, and to examine whether there is a need for large reductions in O/H to reproduce line emission. Section \ref{sec:description} contains a description of the model, and \S\ref{sec:results} describes our modeling results. In §\ref{sec:comparison} we compare our modeling results with constraints obtained with Herschel, and discuss differences between our results and previous models in \S\ref{sec:discussion}. We present our conclusions in Section \ref{sec:conclusions}.

\section{Model description} 
\label{sec:description}

For our analysis, we use data from the models presented in \citet{Ruaud22}, which were
obtained using the disk modeling framework described in \citet{Ruaud19}. We provide a brief summary of the modeling procedure here and we refer to \citet{Ruaud19, Ruaud22} for more details. The models are solved from a minimal set of input physical parameters. Given the stellar properties and the disk gas and dust surface density distributions, the disk physical structure is determined by solving for vertical hydrostatic gas pressure equilibrium, coupled iteratively with gas heating/cooling and chemistry. Many iterations are needed to solve for the self-consistent disk physical structure, and in order to keep the problem tractable, we solve for steady-state chemistry while computing the gas line cooling. Dust grains follow a power-law size distribution ($n_d(a)\propto a^{-3.5}$) and are composed of carbon grains and silicates. The gas/dust mass ratio is assumed to be constant at all radii and $a_\mathrm{min}$ and $a_\mathrm{max}$ at a given $r$ are determined from a coagulation/fragmentation equilibrium approach \citep[for details see][]{Gorti15}. Dust grains settle vertically with a scale height set by the gas turbulence (we adopt a constant value $\alpha = 5\times 10^{-3}$). The density and temperature structure obtained at every $(r,z)$ of both the gas and the dust is then used to solve for time-dependent gas-grain chemistry. For grain chemistry, the ice surface and ice mantle are treated as two separate phases in interaction, and grain surface chemistry includes diffusion of the chemical species on the surface, two-body reactions, photoreactions, and thermal and nonthermal desorption mechanisms. As will be discussed later, water vapor is mainly produced by photodesorption, and the yield measured by laboratory experiments ranges from $10^{-3}$ \citep{CruzDiaz18,Fillion22} to $\sim 3\times 10^{-4}$ and depends on temperature \citep{Fillion22}. We assume a fixed yield of $Y_\mathrm{pd,sH_2O}=10^{-3}$ as our standard value. For photoprocesses, we take into account the absorption by the ions, atoms and molecules of the gas and  the ice and split the UV-visible bands into ten energy bins from 0.74 to 13.6eV, which  were chosen to correspond to dominant gas photoabsorption thresholds. X-ray ionization rates are computed based on the local gas attenuation column\citep{Gorti04}, and cosmic ray ionization is considered to be depth-dependent \citep{Padovani18}.

In total, there are 24 different models where we varied the gas mass, the dust/gas mass ratio and the radial extent of the disk \citep{Ruaud22}. The surface density profile follows $\Sigma(r)=\Sigma_0 (r_c/r) e^{-r/r_c}$ with $r_c = 100$ au and $\Sigma_0$ is  $e \Sigma(r_c)$. The inner disk radius is set to $r_\mathrm{in} = 1$ au and the standard outer radial cut-off for the grid is set to 300 au. For the gas mass, we considered masses in the range $3\times10^{-4} - 10^{-1} M_\odot$  with increments of 0.5 dex. For the dust/gas mass ratio we explored three different ratios of $\epsilon=10^{-1}$, $10^{-2}$, $10^{-3}$. The effect of the radial extent of the disk was explored only for model series with $\epsilon=10^{-2}$ where we consider an additional set of models with a more compact disk set to an outer radius of 100 au. The adopted initial elemental abundances are as in the interstellar medium (ISM) and in particular we use O/H=$3.2\times 10^{-4}$ and C/H=$1.4\times 10^{-4}$. Stellar parameters were kept fixed with $M_* = 1M_\odot$, $R_* = 2.45 R_\odot$, and an X-ray luminosity of $10^{30}$ erg s$^{-1}$. The X-ray spectrum we use is a two-temperature plasma with components at 5 and 10MK. The stellar UV$+$visible spectrum is the TW Hya spectrum taken from the Leiden database \citep{Heays17}, and cross-sections in each UV bin for photo-reactions are computed by integrating over these spectra.  Emission line fluxes are computed using an escape probability method assuming face-on disks (zero inclination). Non-LTE radiative transfer includes collisional and radiative processes, and includes radiation from the dust background, which is iteratively computed with the disk physical and chemical structure. Molecular data are from the Leiden Atomic and Molecular Database (LAMDA). For water excitation, the main collision partner  is H$_2$. In what follows, we assume that the ortho/para ratio of water and H$_2$ are both equal to 3. Our non-LTE calculation has been benchmarked against the radiative transfer code LIME \citep{Brinch2010} and the two methods agree to within a factor $\lesssim 1.5$ for the considered water lines. For details on the chemical network and rates, please see \citet{Ruaud19}. 

\begin{figure*}
    \centering
    \includegraphics[width=1\textwidth]{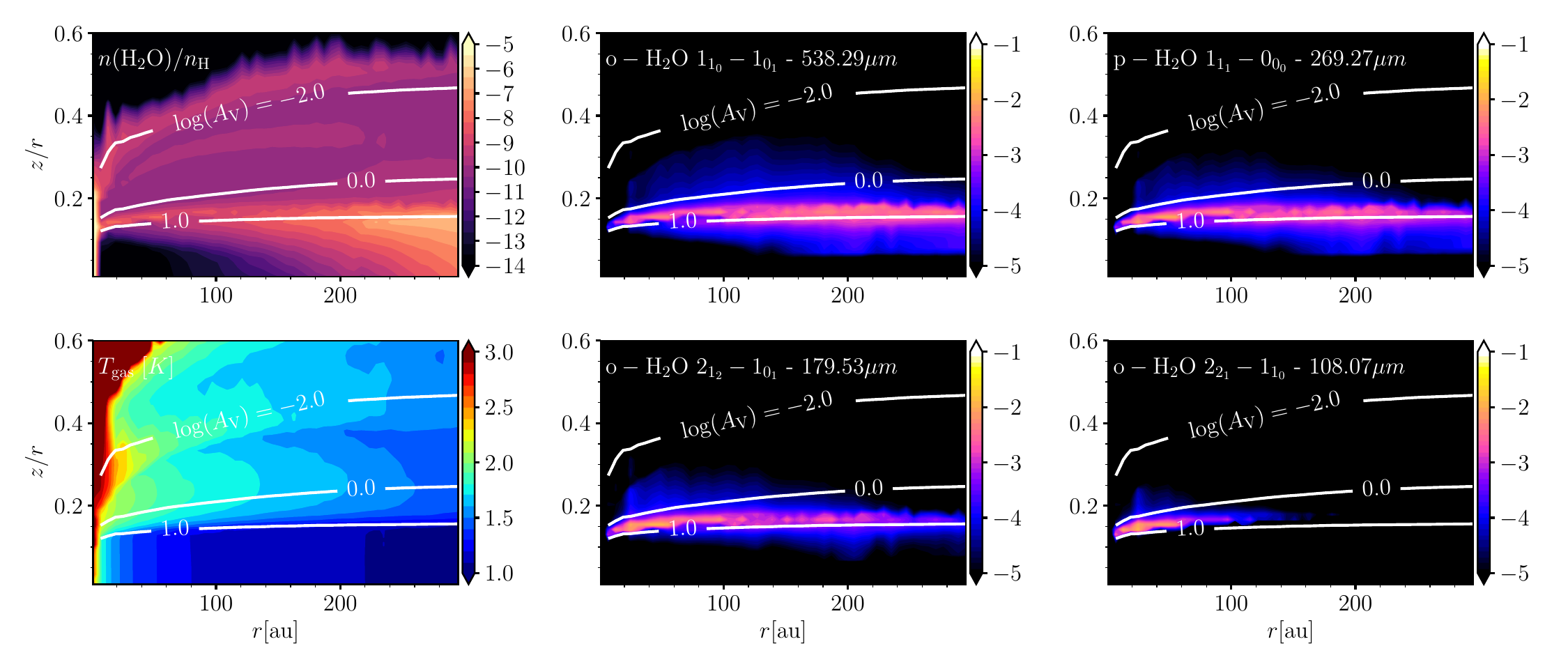} 
    \caption{Computed abundance map of water vapor (top left), gas temperature (bottom left) and predicted line emission map normalized by the total emission of o-H$_2$O $1_{10}-1_{01}$ (at $538.29\mu m$), p-H$_2$O $1_{1_0}-0_{0_0}$ (at $269.27\mu m$), o-H$_2$O $2_{1_2}-1_{0_1}$ (at $179.53\mu m$) and o-H$_2$O $2_{2_1}-1_{1_0}$ (at $108.07\mu m$) for a model disk with $\epsilon=10^{-2}$ and $M_\mathrm{gas}=10^{-2}M_\odot$.}
    \label{fig:water_abun}
\end{figure*}

\begin{figure}
    \centering
    \includegraphics[width=0.45\textwidth]{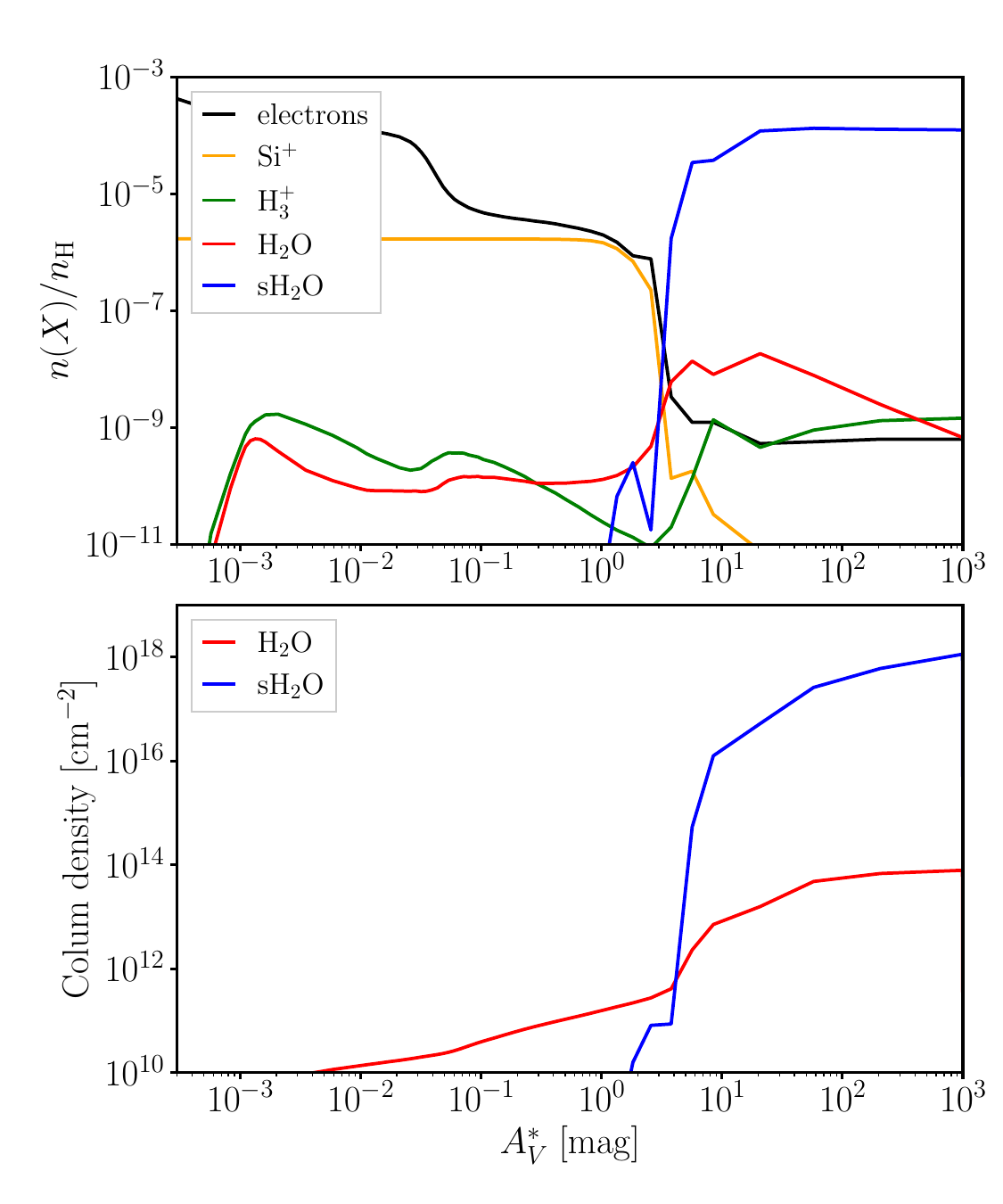} 
    \caption{Top: Computed gas phase abundance of electrons, $\mathrm{Si^+}$, $\mathrm{H_3^+}$, $\mathrm{H_2O}$ and ice phase $\mathrm{sH_2O}$ as a function of the extinction to the star ($A_V^*$) at $r=200$ au. Bottom: Cumulative vertical column densities of water vapor and ice as a function of $A_V^*$ at $r=200$ au. The column density of water vapor rises at a few $A_V^*$ and plateaus right near the vertical snowline. }
    \label{fig:water_cut}
\end{figure}

\section{Results}
\label{sec:results}
\subsection{Cold water distribution and emission in disks}
Figure \ref{fig:water_abun} shows the computed density map of water vapor in the disk as well as 
the gas temperature for a model disk with $\epsilon=10^{-2}$ and $M_\mathrm{gas}=10^{-2}M_\odot$. The calculated line emission maps of the \textit{Herschel}-detected H$_2$O lines at 538.29 (${1_{1_0}-1_{0_1}}$), 260.27 (${1_{1_1}-0_{0_0}}$), 179.53 (${2_{1_1}-1_{0_1}}$) and 108.07 $\mu m$ (${2_{2_1}-1_{1_0}}$) are also shown, normalized to the total emission for each line.  As seen in this figure, water vapor is abundant mostly in a narrow region near the disk midplane at all radii and with a nearly constant abundance of few $\times 10^{-8}$ with respect to $n_\mathrm{H}$ (with peak abundances on the order of $\sim 10^{-7-8}$ in the outer disk). This region marks the interface between the warm molecular layer and the icy disk midplane, and water vapor is mainly produced by X-ray ionization in the inner disk (i.e. at $r\lesssim 75$ au for the model shown in Fig. \ref{fig:water_abun}) and by photodesorption in the outer disk (i.e. at $r\gtrsim 75$ au).  Cold water lines  are seen to originate from a very narrow band near this peak in abundance in Fig.~\ref{fig:water_abun}.

Water emission at 538.29, 269.27 and 179.53$\mu m$ is extended and traces the full radial extent of the disk. The o-H$_2$O $2_{2_1}-1_{1_0}$ emission at 108.07$\mu m$ only extends to $r\sim 75$au, as the temperature of the gas in the emitting region is not high enough in the outer disk (i.e. $T_\mathrm{gas}\sim 20-40$K) to populate the upper-level energy of $E_\mathrm{up}=194$K. Although not as clearly visible as for the  $2_{2_1}-1_{1_0}$ line, other lines shown here also originate from regions that are in general colder than the upper energy-level of the transition. The emission of these other lines is therefore also sensitive to the gas temperature (i.e. the emission is $\propto \exp(-E_\mathrm{up}/T_\mathrm{gas})$.

The main processes that lead to the water emission layer are described by examining in detail a vertical cut through the disk. We provide simple analytical estimates to show the dependency of water vapor abundance on the different disk and chemical parameters. For simplicity, we discuss the photo-reactions in terms of the typical measure $G_0$ for the entire UV field from $6-13.6$eV. Note that the full disk models use several FUV bins, but the main chemical pathways can be understood by considering the total UV field. 
Figure \ref{fig:water_cut} illustrates the dependencies we set out to explain in what follows; we mainly focus on explaining the column density of water shown in the lower panel. The upper panel of Fig.~\ref{fig:water_cut} also shows the vertical abundance profile of electrons, $\mathrm{Si^+}, \mathrm{H_3^+}$, $\mathrm{H_2O}$ and $\mathrm{sH_2O}$ ('s' here denotes ice chemical species) as a function of the extinction toward the star ($A_V^*$) at $r=200$ au. UV photoprocesses set the abundance of water vapor in the outer disk, and therefore we first discuss the attenuation of the FUV photon flux into the disk.

\subsubsection{Local attenuated FUV photon flux}
\label{sec:fuvatten}
The local FUV flux in the disk at $(r,z)$ has two components, the attenuated FUV flux from the star and the attenuated FUV flux from the ISM. We assume an unattenuated ISM FUV field $G_0^\mathrm{ISM}=1$, while the stellar field from the adopted UV spectrum yields $G_0^* \sim 10^7$ in Habing units at 1 au. 
These photons are absorbed by dust, atoms and molecules from the gas and the ice, including water ice \citep[see][]{Ruaud19}.
At the disk surface, the FUV photon flux is dominated by the star ($G_0^*\sim 250$ at $r=200$ au) and opacity from the dust and gas species along the line of sight to the star contribute to the attenuation which increases with decreasing $z$. Deeper into the disk, the stellar FUV photon flux drops rapidly with increasing $A_V^*$ (attenuation toward the star in the visual, which we use a proxy for $\tau_{UV}$ in what follows) as it gets attenuated. The local UV flux eventually becomes low enough that the rate at which oxygen atoms accrete at the surface of grains becomes larger than the rate at which water ice is photodesorbed from the surface of grains (at $A_V^* \sim 3$ mag and $r=200$au in the model with $\epsilon=10^{-2}$ and $M_\mathrm{gas}=10^{-2}M_\odot$ shown in Fig. \ref{fig:water_cut}).
From this point on, water ice rapidly accumulates at the surface of the grains and the abundance of water ice rises exponentially with depth until most of the available oxygen is locked into ices.
Somewhere in this region, the ISM UV field which is less attenuated may begin to dominate the local UV flux.
When \nice\ becomes higher than $\sim 2\times 10^{17}$ cm$^{-2}$, the opacity of water ice is on the order of $\tau_\mathrm{sH_2O} = \sigma_\mathrm{sH_2O} N_\mathrm{sH_2O}\sim 1$ (assuming an absorption cross section for water ice $\sigma_\mathrm{sH_2O} \sim 5\times 10^{-18}$ cm$^2$). At $r=200$au, this typically occurs at $A_V^*\gtrsim 10$ mag, where the visual extinction in the vertical direction due to dust is 
$\gtrsim 0.1$ mag.  Here, even the ambient ISM field in the vertical direction is attenuated by the built-up water ice column density. 

To summarize, the FUV flux from the star at a given radius sets the location of the vertical water snowline, around $A_V^*\sim 3$ mag at $r=200$ au. With increasing vertical depth, the stellar FUV flux drops rapidly, and by $A_V^*\gtrsim 3$ mag the interstellar FUV flux begins to dominate the local UV field and may cause photodesorption. At about $A_V^*\sim 10$ mag water ice begins to completely shield itself. Photodesorbed gas phase water, therefore, is most abundant in between the region where 
$3\lesssim A_V^* \lesssim 10$ mag, as seen in Figure~\ref{fig:water_abun}.

\subsubsection{Formation of water vapor at the disk surface}
\label{sec:gas_chem}
In this section, we focus on the chemistry in regions with $A_V^* \lesssim 3-5$ mag, where gas-phase processes dominate the formation of water vapor. In the inner disk, i.e. at $r\lesssim 100$ au, water formed in the gas phase can significantly contribute to line emission (further details are in Section \ref{sec:sensitivity}). As seen in Fig. \ref{fig:water_cut}, the water vapor abundance at the disk surface is on the order of few $\times 10^{-10}$ in the region between $10^{-3}\lesssim A_V^* \lesssim 1$ mag. Here, water vapor formation is dominated by ion-chemistry. It starts with 
the ionization of H$_2$ by X-rays and cosmic-rays to form $\mathrm{H_3^+}$, and then the reaction of O with $\mathrm{H_3^+}$ to form OH$^+$. OH$^+$ successively reacts with H$_2$ to form H$_2$O$^+$ and H$_3$O$^+$. H$_3$O$^+$ then recombines with electrons to form water and OH with a branching ratio of $1/4$ and $3/4$ respectively. Destruction of water vapor in this region is dominated by photodissociation by stellar photons. Just above the disk midplane reaction with Si$^+$ participates to the destruction of water vapor until most of the silicon (we assume [Si/H]=$1.7\times 10^{-6}$) depletes at the surface of the grains. Near the water snowline reaction with H$_3^+$ also contributes to the destruction of water vapor. There is a decrease here in the abundance of $\mathrm{H_3^+}$ which is being primarily destroyed by reactions with CO. When CO freezes near the water snowline, destruction of $\mathrm{H_3^+}$   becomes dominated by reaction with electrons and $\mathrm{H_3^+}$  abundance increases again. Just above the vertical water snowline (i.e. at $A_V\sim3$ mag in the model with $\epsilon=10^{-2}$ and $M_\mathrm{gas}=10^{-2}M_\odot$), where vapor production by photodesorption starts to gain in significance, the abundance of water vapor formed by ion chemistry is found to increase to few $\times 10^{-9}$.

Due to the rapid decrease in density with increasing height, water formed by ion chemistry at the disk surface only marginally contributes to the total vertical column density of water vapor (\nwat) in the outer disk once vapor production by photodesorption is taken into account. However, this purely gas-phase route provides a floor to the total vertical column of water vapor; for example, this is a few $\times 10^{12}$ cm$^{-2}$ at $r=200$ au in the model with $\epsilon=10^{-2}$ and $M_\mathrm{gas}=10^{-2}M_\odot$.

\subsubsection{Simple analytical estimate of the water vapor produced by photodesorption}
\label{sec:grain_chem}
Having described the attenuation of the FUV photon flux into the disk (\S \ref{sec:fuvatten}), we can derive a simple analytical estimate of the distribution of water in the disk. We arbitrarily pick a radius of 200au to illustrate the main processes at work. As described earlier, the location of the photodesorption layer (i.e. where water vapor peaks in abundance at $r=200$ au) is set by the location of the vertical water snowline (at $A_V^*\sim 3$ mag) and the vertical column of water ice needed to attenuate the FUV photon flux (at $A_V^*\sim 30$ mag). At this location, water abundance is $\sim 10^{-8}$ at $r=200$ au (see Fig. \ref{fig:water_cut}). We note that depending on the adopted photodesorption yield, chemical desorption may also contribute to the formation of gas phase water. Chemical desorption starts to compete with photodesorption when the photodesorption yield drops to $Y_\mathrm{pd,sH_2O} \lesssim 10^{-4}$. This is discussed in more detail in Sections \ref{sec:var_phodes_chemdes} and \ref{sec:col_dens}.
The basic assumption behind the chemical desorption process is that for exothermic reactions on ices, part of the energy released during the reaction can be used to desorb the product to the gas phase. In our model we assume an efficiency for this process of $\sim 1\%$ \citep[see][for more details on this process and its implementation into the model]{Ruaud19}. In regions where water vapor emits, we find that photodesorption is the main mechanism producing water vapor and therefore neglect the effect of chemical desorption in the following (this will be discussed in more detail in Sec. \ref{sec:col_dens}). By neglecting chemical desorption,  a simple expression can be derived to understand the production of water by photodesorption in disks in an approach similar to \citet{Hollenbach09}. In order to estimate water vapor abundance, we first determine where ice production begins. The abundance of water ice as a function of $z$ can be found by equating the formation rate (oxygen accreting at the surface of grains) to the destruction rate (photodesorption to the gas phase):

\begin{equation}
k_\mathrm{acc,\mathrm{O}}\ n(\mathrm{O}) = k_\mathrm{pd,\mathrm{sH_2O}}\ n(\mathrm{sH_2O})
\end{equation}
Assuming all oxygen is in the form of gas phase oxygen and water ice,  the conservation equation for oxygen $n_\mathrm{O} = n(\mathrm{O})+n(\mathrm{sH_2O})$, leads to
\begin{equation}
\label{eq:sh2o}
X(\mathrm{sH_2O}) = \frac{k_\mathrm{acc,\mathrm{O}}}{k_\mathrm{acc,\mathrm{O}} + k_\mathrm{pd,\mathrm{sH_2O}}}X_\mathrm{O},
\end{equation}
where $n$ denotes density and $X$ denotes abundance, with $X_O$ being the total elemental abundance of oxygen. We next estimate $X(\mathrm{H_2O})$  by equating its formation rate (photodesorption of water ice)  to its gas-phase destruction  rate. 
The rate at which water ice is photodesorbed is given by
\begin{equation}
\label{eq:evap_water}
\begin{split}
R_\mathrm{pd,sH_2O} &= k_\mathrm{pd,sH_2O} \ n(\mathrm{sH_2O}) \\
 &= \frac{ F_0 G_\mathrm{UV} Y_\mathrm{pd,sH_2O} \ \langle \sigma_\mathrm{d} n_\mathrm{d} \rangle }{n_{ice}} n(\mathrm{sH_2O}) \\
&= F_0 G_\mathrm{UV} Y_\mathrm{pd,sH_2O} \ \langle \sigma_\mathrm{d} n_\mathrm{d} \rangle f_\mathrm{sH_2O} 
\end{split}
\end{equation}
where $f_\mathrm{s,sH_2O} = n(\mathrm{sH_2O})/n_{ice}$ 
is the fraction of ice (in number) present at the surface that is in water, $F_0$ is the energy flux to photon number conversion factor in photons cm$^{-2}$ s$^{-1}$. Here,  $G_\mathrm{UV}(r,z)$ denotes the local UV flux, both from the star and the ISM, including the attenuation factors due to dust, atoms and molecules (from the gas and the ice) in both directions (i.e vertical and from the star).  Eq.~\ref{eq:evap_water} is only valid in the multilayer regime (i.e. after the first monolayer of ice has formed). In the case where only water ice chemistry is considered, $f_\mathrm{sH_2O}=1$. In our models however, $f_\mathrm{sH_2O}$ is lower than 1 in regions with $3\lesssim A_V^* \lesssim 10$ mag due to the efficient formation of sCO$_2$ ice induced by photoprocessing of water ice \citep[through the reaction of sOH with sCO, see][]{Ruaud19,Ruaud22}. At $r=200$ au, $f_\mathrm{sH_2O} \sim 5\times10^{-2}$ at $A_V^* \sim 3$ mag (where ices start to form) and rapidly increases to $f_\mathrm{sH_2O} \sim 0.5$ at $A_V^* = 10$ mag. 

For the destruction rate of photodesorbed water vapor, we have two main routes: (i) photodissociation by stellar and interstellar photons and (ii) destruction by reaction with H$_3^+$. 
The rate of photodissociation of water vapor is given by:
\begin{equation}
\label{eq:diss_water}
R_\mathrm{diss,H_2O} = k_0 G_\mathrm{UV}  n(\mathrm{H_2O})
\end{equation}
where $k_0$ is the unshielded water photodissociation rate when $G_0 = 1$ and is approximately $k_0 \sim 10^{-9}$s$^{-1}$ for H$_2$O. 
If we further ignore destruction of water vapor by reaction with H$_3^+$, we can find the abundance at which gas-phase water saturates:
\begin{equation}
\label{eq:xplh2o}
X_\mathrm{sat}(\mathrm{H_2O}) = \frac{k_\mathrm{pd,\mathrm{sH_2O}}}{k_\mathrm{diss,H_2O}}X(\mathrm{sH_2O})
\end{equation}
Using Eqs. \ref{eq:evap_water} and \ref{eq:diss_water} and introducing the cross-sectional grain area per H atom $\sigma_\mathrm{H} = \langle \sigma_\mathrm{d} n_\mathrm{d} \rangle / n_\mathrm{H}$, we get:
\begin{eqnarray}
\label{eq:h2o_sat}
X_\mathrm{sat}(\mathrm{H_2O}) &=& \frac{Y_\mathrm{pd,sH_2O} f_\mathrm{sH_2O} \sigma_\mathrm{H} F_0}{k_0} \\
&\sim& 10^{-8} f_\mathrm{sH_2O} \Bigg( \frac{Y_\mathrm{pd,sH_2O}}{1\times10^{-3}}\Bigg) \Bigg( \frac{\sigma_\mathrm{H}}{1\times10^{-22}}\Bigg)
\end{eqnarray}
Eq.~\ref{eq:h2o_sat} shows that after the first monolayer of ice has formed (i.e. at $A_V^*\gtrsim 2-3$ mag from our models), the abundance of gas phase water saturates to a value that is independent of the strength of the FUV flux and only depends by $f_\mathrm{sH_2O}$, $Y_\mathrm{pd,sH_2O}$ and $\sigma_H$. The lack of dependence on FUV photon flux is due to the fact that both the formation and destruction rates of water vapor linearly depend on G$_0$ which cancel out at equilibrium, if only photoprocesses are considered \citep{dominik2005,Hollenbach09}. 
Including additional destruction pathways will lower the abundance of water vapor from its saturation value. For example, accounting for the next dominant destruction route induced by reaction with $\mathrm{H_3^+}$, the abundance of water in the gas can be expressed as a function of $X_\mathrm{sat}(\mathrm{H_2O})$ by:
\begin{equation}
\label{eq:xh2o}
X(\mathrm{H_2O}) = \Big( 1+\frac{0.75 k_1 n(\mathrm{H_3^+})}{k_0 G_\mathrm{UV}} \Big)^{-1} X_\mathrm{sat}(\mathrm{H_2O})
\end{equation}
where $k_1$ is the rate coefficient for the reaction between H$_2$O and H$_3^+$ leading to the formation of H$_3$O$^+$(the factor 0.75 arises because H$_3$O$^+$ recombines 1/4 of the time to reform water) and is $\propto T^{-1/2}$ . This expression is similar to the expression given in \citet{Hollenbach09} with the difference that it includes two directions (to the star and vertical) and includes shielding by water ice, both of which are folded into the local UV field $G_\mathrm{UV}$. 

As seen from Eq.~\ref{eq:xh2o},
the abundance of water vapor will begin to decrease when the second term in the denominator starts exceeding unity, i.e., when the $\mathrm{H_3^+}$ destruction route becomes higher than the local UV photodissociation rate of water. From the models, this happens when the vertical column of water ice exceeds $\sim 10^{17}$ cm$^{-2}$, and can shield vapor from photodissociation.  As a consequence, the gas phase abundance of water is close to $X_\mathrm{sat}(\mathrm{H_2O})$ as long as the vertical column of water ice is below $\sim 10^{17}$ cm$^{-2}$, which happens deep in the disk at at $A_V^*\gtrsim 30$ mag (Fig.~\ref{fig:water_cut}).

\begin{figure}
    \centering
    \includegraphics[width=0.45\textwidth]{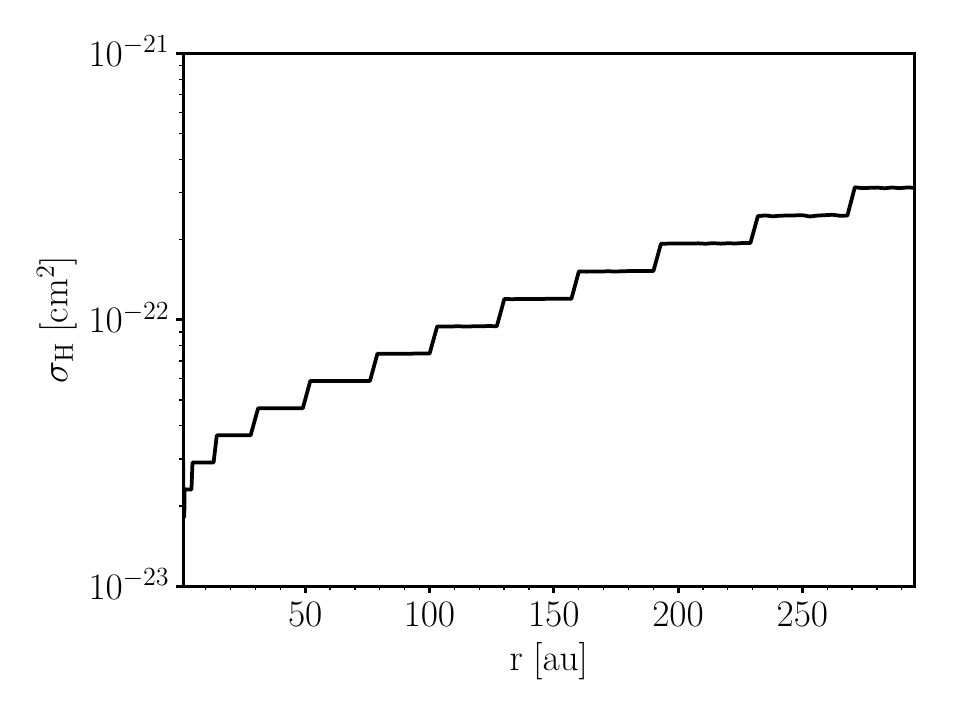}
    \caption{Midplane dust cross sectional area per H atom ($\sigma_\mathrm{H}$) as a function of the disk radius for the disk model with $\epsilon = 10^{-2}$ and $M_\mathrm{gas} = 10^{-2} M_\odot$.}
    \label{fig:sigmaH}
\end{figure}

In the regime where $X(\mathrm{H_2O}) \sim X_\mathrm{sat}(\mathrm{H_2O})$, the only model parameters that affect the abundance of gas phase water are the photodesorption yield ($Y_\mathrm{pd,sH_2O}$) and the dust cross-sectional area per H atom ($\sigma_\mathrm{H}$). The dust cross-sectional area per H atom is proportional to $\propto (a_\mathrm{min}a_\mathrm{max})^{-0.5}$ and $a_\mathrm{max}$ is proportional to the local gas density \citep{birnstiel2012}. As a result, an increase in the local gas disk density leads to larger $a_\mathrm{max}$ and therefore lower $\sigma_\mathrm{H}$.  Our assumption of constant gas/dust mass ratio means that in the outer disk, as the surface density falls, the maximum grain size (which is $\propto \Sigma$) decreases and the dust is distributed in smaller grains. $\sigma_\mathrm{H}$ therefore increases with radius (see Fig. \ref{fig:sigmaH}), by a factor of a few through the entire disk. This increases the water column density in the outer disk by a similar factor  because $X(\mathrm{H_2O}) \propto \sigma_\mathrm{H}$. Therefore, there is only a weak dependence of the water distribution on  $\sigma_\mathrm{H}$ and the disk dust properties. 

\subsection{Sensitivity to cosmic ray and X-ray ionization rates, $Y_\mathrm{pd,sH_2O}$ and chemical desorption}
\label{sec:sensitivity}
Since our analysis above indicates that $X(\mathrm{H_2O})$ is sensitive to the cosmic ray and X-ray ionization rates (\S \ref{sec:gas_chem}), as well as to the somewhat uncertain photodesorption yield (which could be closer to few $10^{-4}$ and vary with temperature \citep{Fillion22}) and chemical desorption, we study the effect of varying these parameters on the disk model with $\epsilon=10^{-2}$, $M_\mathrm{gas}=10^{-2}$ and $Y_\mathrm{pd,sH_2O}=10^{-3}$ (labeled as the fiducial model in the following).  In a first series of models, we retain chemical desorption and the photodesorption yield at $Y_\mathrm{pd,sH_2O}=10^{-3}$, and first vary the ionization rates. For the cosmic ray ionization rate, we consider a variation of the fiducial depth-dependent model where we use a low, constant, ionization rate of $\zeta_{H_2}=10^{-19}$ s$^{-1}$ \citep[e.g.][]{Cleeves14,Cleeves15}. For comparison, $\zeta_{H_2}\sim10^{-17}$ s$^{-1}$ in the midplane for $\Sigma_{gas}=0.053$ using our depth dependent expression \citep[i.e. for which we use][]{Padovani18}. In a second model, we use $\zeta_{H_2}=10^{-19}$ s$^{-1}$ and also turn off X-ray ionization. 
For the photodesorption yield and chemical desorption, we consider three variations of the fiducial model. In a first model we turn off photodesorption and chemical desorption (i.e. we consider that desorption from grains can only be thermal). In a second model we use a photodesorption yield that is $10 \times$ lower than the fiducial case (i.e. $Y_\mathrm{pd,sH_2O}=10^{-3}$) and where chemical desorption is turned off. Finally, in a third model we also use a photodesorption yield of $Y_\mathrm{pd,sH_2O}=10^{-4}$ and turn off chemical desorption.

\begin{figure}
    \centering
    \includegraphics[width=0.45\textwidth]{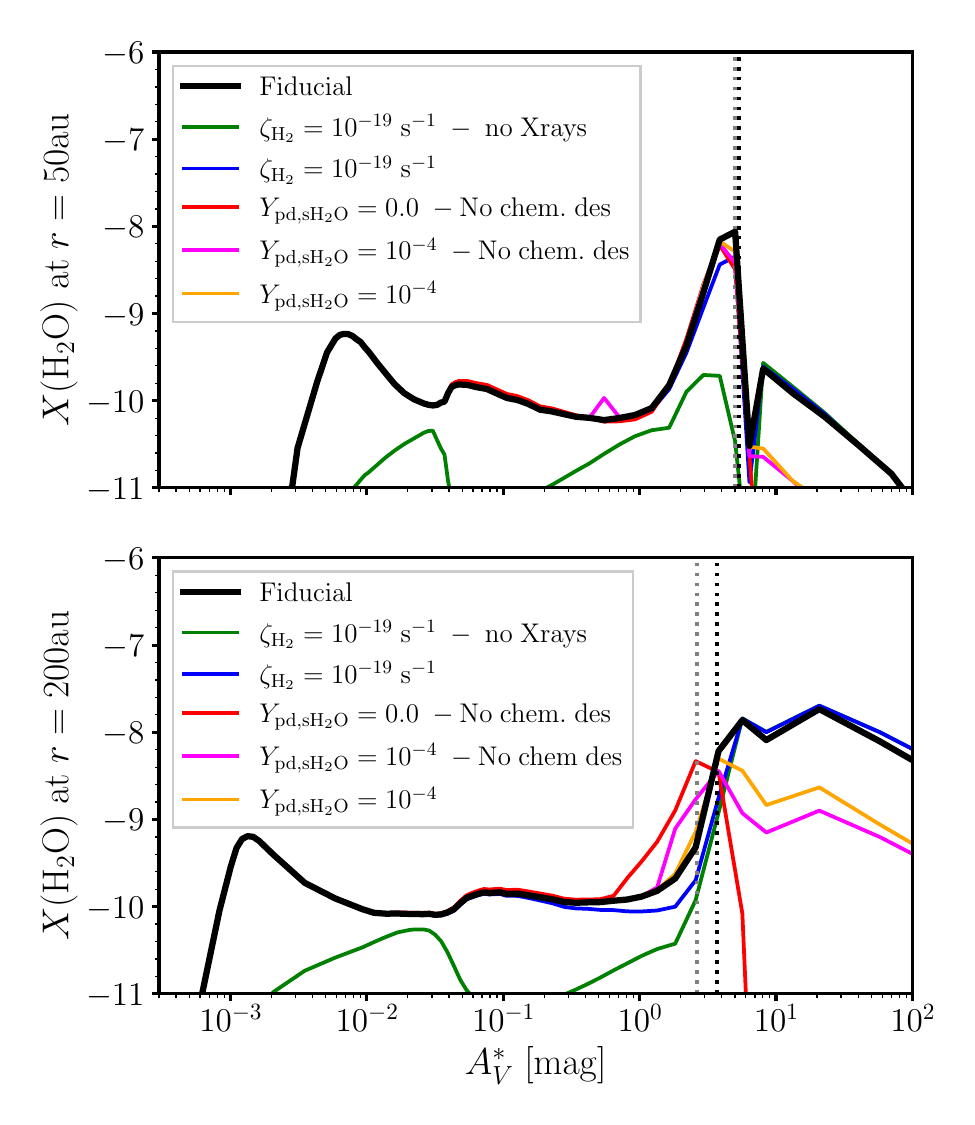}
    \caption{Vertical water vapor abundance profile at $r=50$ au and $r=200$ au for the disk model with $\epsilon=10^{-2}$, $M_\mathrm{gas}=10^{-2}$ and $Y_\mathrm{pd,sH_2O}=10^{-3}$, and five variations of the same model. In two models (blue and green lines) we study the impact of decreasing the cosmic ray and X-ray ionization rate. In the last three models (red, magenta and orange) we study the impact of varying the photodesorption yield and and chemical desorption. In all these models the location of the vertical water snowline is  identical and is shown here by the vertical gray dotted line. We define the water snowline as the location where $50\%$ of the oxygen is on the form of ice at the surface of grains. The vertical black dotted line denotes the location where $90\%$ of the oxygen is on the form of ice.
    }
    \label{fig:additional_models}
\end{figure}

Fig. \ref{fig:additional_models} shows the vertical water abundance profile for these six models at $r\sim 50$ au (upper panel) and $r\sim 200$ au (lower panel). These are two representative radial cuts, the full radial behavior is discussed later. While the surface abundance profiles are similar at both radii, \wvap\ below the snowline ($A_V^*\gtrsim 3$ mag) 
is almost two orders of magnitude lower at $r= 50$ au than at $r= 200$ au. This is caused by: (1) decreased photodesorption due to a lower value of $\sigma_\mathrm{H}$ closer to the star (i.e. $\sigma_\mathrm{H}$ in the midplane is a factor of $\sim 4$ lower at $r=50$au compared to $r=200$au, see Fig. \ref{fig:sigmaH}), which decreases the $X_\mathrm{sat}(\mathrm{H_2O})$ in this layer by the same amount (see Eq. \ref{eq:h2o_sat}), and (2) the higher gas density in the inner disk resulting in a higher attenuation of the FUV photon flux by atoms and molecules from the gas and the ice at a given $A_V^*$, which therefore reduces $X(\mathrm{H_2O})$ in the disk midplane (see Eq. \ref{eq:xh2o}).

\subsubsection{Dependence of gas-phase formation of water vapor on cosmic ray and X-ray ionization rates}
As seen in Fig. \ref{fig:additional_models}, cosmic rays and X-rays only impact the abundance of water vapor in the disk atmosphere (i.e. at $A_V^*\lesssim 3-5$ mag), where gas-phase processes dominate its formation (see \S \ref{sec:gas_chem}). 
Lowering $\zeta_{H_2}$ alone (blue vs. black lines in Fig.~\ref{fig:additional_models}) has little effect on the computed \wvap\ compared to the fiducial case. However, in the case where X-rays are also turned off while $\zeta_{H_2}$  is kept low, \wvap\ is greatly reduced in the disk atmosphere (green vs. black lines). This is more so in the inner disk where gas-phase ion-molecule chemistry is the main contributor to the total \nwat. Here, \wvap\  is reduced by $\sim 2$ dex at $r=50$ au and $A_V^*\sim 3$ mag, because a low $\zeta_{H_2}$ with no X-rays leads to a correspondingly low  formation rate of H$_3^+$, the key ion leading to the gas-phase formation of H$_2$O (see Section \ref{sec:gas_chem}). In the fiducial model, X-rays dominate ionization at $r=50$ au and drive the increase in \wvap. 
For this model with low  $\zeta_{H_2}$ and no X-ray ionization, the change in the resulting luminosity for water lines originating from the inner disk (i.e. at $r\lesssim 100$ au) like the o-H$_2$O lines at $108.07\mu$m and at $179.53\mu$m, is significant (to be discussed later). 

At  $r=200$ au, there is again a strong decrease in \wvap\ for this model (low $\zeta_{H_2}$, no X-rays), more so at the surface ($A_V^* < 3$ mag) compared to nearer the snowline. 
As discussed earlier in \S ~\ref{sec:gas_chem}, ion-molecule chemistry only provides a floor to the total water column density, and as seen in Fig. \ref{fig:additional_models}, reducing the cosmic ray and X-ray ionization does not impact the computed midplane water vapor abundance. At $r=200$ au, photodesorption is the main contributor to the total water column density and therefore the total column of water remains largely unaffected with change in ionization rates.

\subsubsection{Dependence of photodesorbed water on $Y_\mathrm{pd,sH_2O}$ and chemical desorption}
\label{sec:var_phodes_chemdes}
Varying the photodesorption yield and chemical desorption affect the abundance of water vapor only in regions with $A_V^*\gtrsim 1$ mag. In the case where both photodesorption and chemical desorption are turned off (i.e. $Y_\mathrm{pd,sH_2O}=0$ and no chemical desorption), the abundance of \wvap\ computed in the region with $1\lesssim A_V^*\lesssim 4$ mag and $r=200$ au increases as compared to the fiducial model (red vs. black lines). This is explained by the fact that in this model the abundance of Si$^+$, a major destruction channel of water in the disk atmosphere (see Section \ref{sec:gas_chem}), is decreased by several orders of magnitude in this region. This decreased abundance of Si$^+$ is explained by the fact that in the absence of photodesorption and chemical desorption, silicon freezes-out higher up in the disk due to its relatively high binding energy with the surface of grains. The lower abundance of Si$^+$ in the region with $1\lesssim A_V^*\lesssim 4$ mag lowers the destruction of water vapor compared to the fiducial case and therefore its abundance increases. For this model, formation of water vapor close to the vertical water snowline  (i.e. at $A_V^*\sim 3-4$ mag) is set by gas-phase chemistry. The efficiency of this formation route decreases as oxygen becomes locked into water ice as $z$ decreases (i.e. at $A_V^*\sim 4$ mag, where $90\%$ of the oxygen is on the form of ices at the surface of grains).
As seen in Fig. \ref{fig:additional_models}, decreasing the photodesorption yield from $Y_\mathrm{pd,sH_2O}=10^{-3}$ to $Y_\mathrm{pd,sH_2O}=10^{-4}$ (i.e. black vs. magenta lines) decreases the abundance of water vapor by a factor of $\sim 10$, as predicted from our simple estimates. However, the change in resulting luminosity is not significant as we show below. Turning off chemical desorption but keeping a photodesorption yield of $Y_\mathrm{pd,sH_2O}=10^{-4}$ (magenta curves) only marginally impacts the computed abundance of water vapor in the disk midplane as compared to the case where chemical desorption is turned on and the photodesorption yield is $Y_\mathrm{pd,sH_2O}=10^{-4}$ (orange curves) (i.e. \wvap\ in the midplane decreases by a factor $\lesssim 2-3$ at a given $A_V^*$ and at $r=200$ au). This shows that chemical desorption starts to compete with photodesorption for producing water vapor in the disk midplane when $Y_\mathrm{pd,sH_2O}\lesssim 10^{-4}$. For higher photodesorpion yields, its formation is completely set by photodesorption.

\begin{figure}
    \centering
    \includegraphics[width=0.45\textwidth]{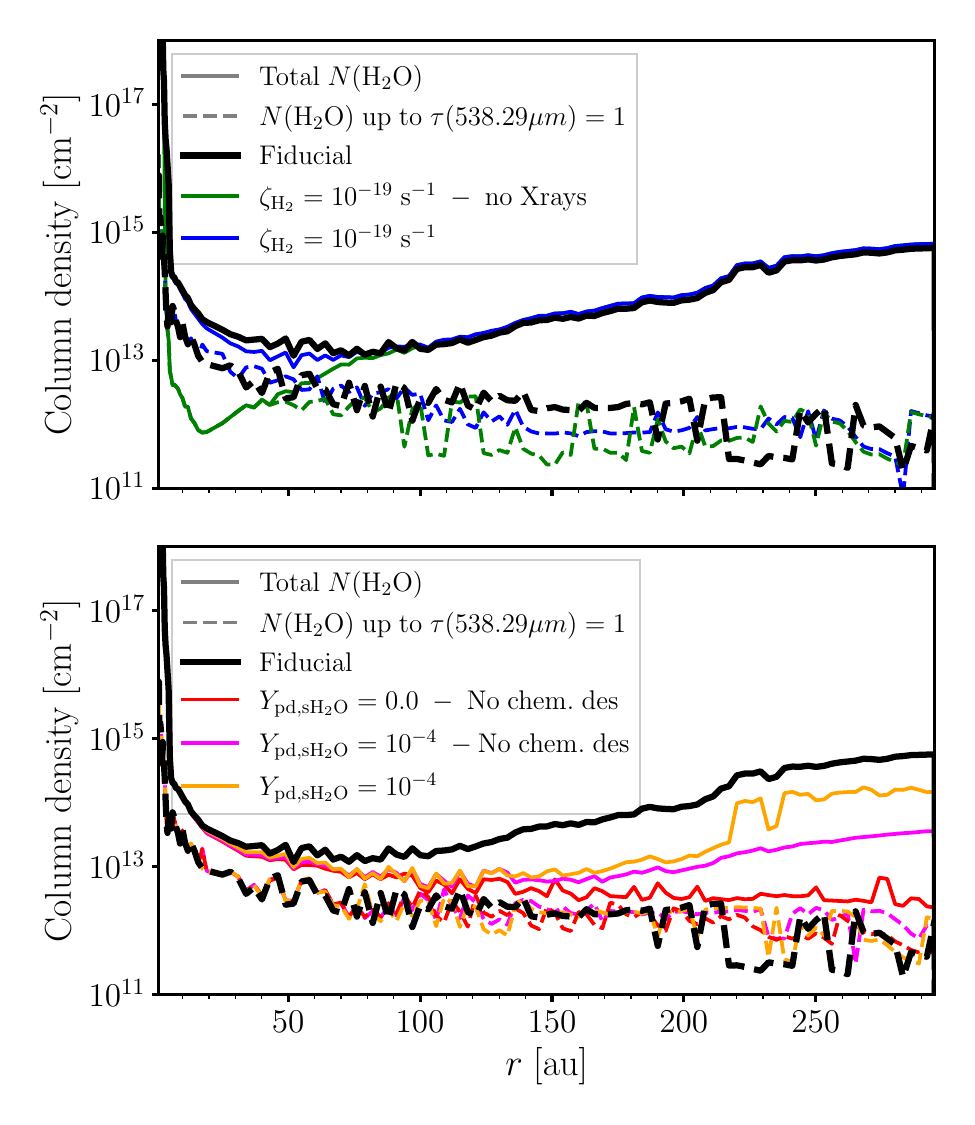}
    \caption{Vertical column density of water vapor as a function of $r$ for the disk model with $\epsilon=10^{-2}$, $M_\mathrm{gas}=10^{-2}$ and $Y_\mathrm{pd,sH_2O}=10^{-3}$, and four variations of the same model. The top panel shows a comparison of fiducial model with models were we decrease the cosmic ray and X-ray ionization rate. The bottom panel shows a comparison of the fiducial model with models were we study the effect of varying the photodesorption yield and chemical desorption.}
    \label{fig:cd_additional_models}
\end{figure}

\subsubsection{Impact of variations on the vertical column density of water vapor and line  luminosity}
\label{sec:col_dens}

Figure \ref{fig:cd_additional_models} shows the computed \nwat\ as a function of $r$ for the fiducial model and the different variations of this model. Solid lines show the total \nwat\  while dashed lines show \nwat\ integrated up to unit optical depth in the o-H$_2$O ground-state line, $\tau (538.29\mu m) = 1$. 

Decreasing the cosmic ray ionization rate does not significantly affect the computed \nwat\ anywhere in the disk, as discussed previously (compare blue and black lines in Fig.~\ref{fig:cd_additional_models}). 
\nwat is more sensitive to X-ray ionization, especially in the inner disk (green vs. black lines in Fig. \ref{fig:cd_additional_models}) where it can decrease by up to a factor $\sim 10$ at $r\sim 20$ au when X-ray ionization is turned off.

Varying the photodesorption yield and switching on/off chemical desorption mainly impacts the outer disk (see \S 3.2.2). When there is no desorption into the gas phase at all (red line), \nwat\ is roughly constant for $r\gtrsim 75$au and on the order few $10^{12}$ cm$^{-2}$, with all water being formed by gas-phase processes and the vertical column density of water being limited by the location of the vertical water snow-line. When the photodesorption yield is decreased by a factor of 10 compared to the fiducial model and chemical desorption is turned off (magenta line vs black line), then \nwat\ in the outer disk regions (i.e. where photodesorption dominates) is decreased by a factor $\sim 10$ as expected from Eq.~\ref{eq:h2o_sat}. In models with $Y_\mathrm{pd,sH_2O}=10^{-4}$, taking into account chemical desorption (orange line), causes an abrupt rise in the total \nwat\ at $\sim 220$ au from $\sim 10^{13}$ cm$^{-2}$ to $\sim 10^{14}$ cm$^{-2}$ compared to the case where chemical desorption is turned off (magenta line). This is explained by the fact that chemical desorption is more efficient for cold dust present in the outer disk midplane; i.e. a longer residence timescale of H atoms on grain surfaces  promotes water ice formation and therefore chemical desorption. 

\begin{figure}
    \centering
    \includegraphics[width=0.45\textwidth]{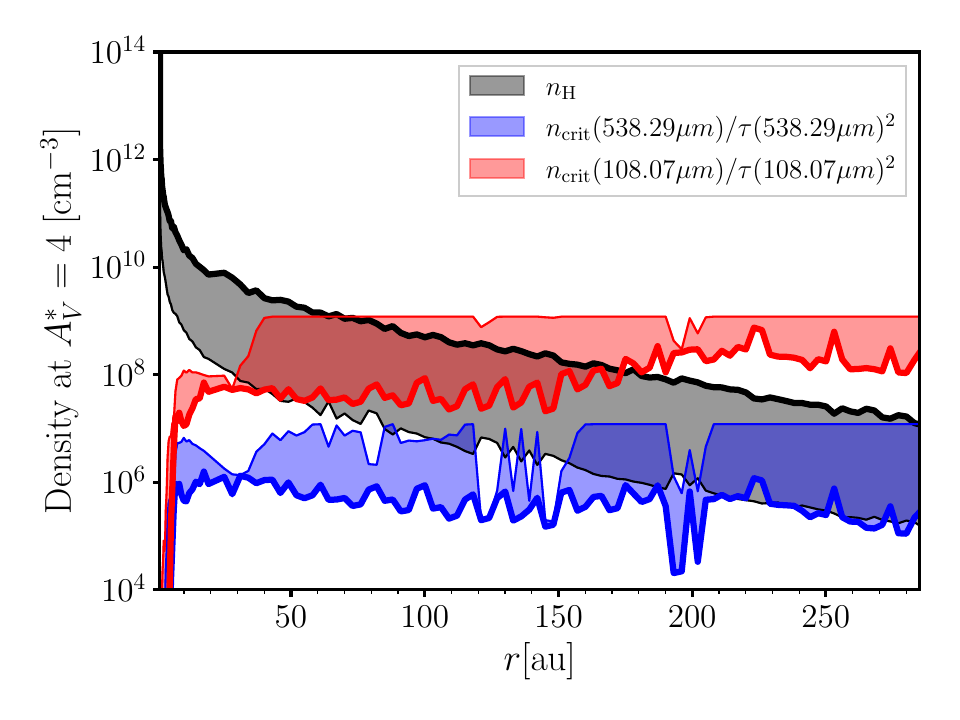}
    \caption{Range of gas density and critical densities weighted by $\tau^2$ when the line is optically thick for the 538.29$\mu m$ and 108.07$\mu m$ lines computed for all the models with $\epsilon=10^{-2}$ and at $A_V^* = 4$ mag as a function of the disk radius. Bold lines enveloping these ranges are for the most massive disk considered in this study (i.e. $M_\mathrm{gas}=10^{-1} M_\odot$ and $\epsilon=10^{-2}$) while thinner lines are for the least massive disk (i.e. $M_\mathrm{gas}=3\times 10^{-4} M_\odot$ and $\epsilon=10^{-2}$).}
    \label{fig:critical_densities}
\end{figure}

However, these changes in the vertical column density of water vapor do not significantly affect the predicted line luminosities because the water lines are in fact optically thick. As seen in Fig.\ref{fig:cd_additional_models}, changes in ionization (with the exception of the low $\zeta_{\mathrm{H_2}}$, no-X-rays model) and desorption mechanisms and yields mainly affect the \nwat\ in the outer disk ($r\gtrsim 100$ au) where the columns correspond to optical depths ranging from 10 to a few 100.  
The high critical densities (defined as the ratio of all transition probabilities to the net collisional rate out of the upper level of a transition) of the water line emission, could result in the lines being effectively thin in which case the emission would be proportional to the column density \citep[see e.g.][]{Linke77,Snell00,tielensbook}. For most of the range of parameters in the parameter survey, however, the densities in the disk at the snow line ( $A_V^* \sim 3-4$ mag) where line emission peaks are still relatively high. Figure ~\ref{fig:critical_densities} shows the density at the snow line as a function of disk radius for the $\epsilon=10^{-2}$ models, compared to $n_{crit}/\tau^2$ for the 
538.29$\mu m$ and 108.07$\mu m$ lines.\footnote{One factor of $\tau$ arises from the escape probability, and the other from line trapping, see \citet{tielensbook} for example.}  Except for very low mass disks, both lines originate from regions with higher collider densities (see Fig. \ref{fig:water_abun} for the spatial distribution of these emission lines).
Therefore, line emission is in general not affected by the changes in \nwat\ as a result of the various parameter explorations discussed earlier, because most of the vertical column is observationally inaccessible. 
For example, for the  o-H$_2$O $1_{1_0}-1_{0_1}$ line at 538.29$\mu m$, although \nwat\ decreases by a factor of $\sim 2-10$ depending on radius, the total luminosity only goes down by a factor of $\sim 1.1$ when  we use $Y_\mathrm{pd,sH_2O}=10^{-4}$ instead of $Y_\mathrm{pd,sH_2O}=10^{-3}$. 
The highest differences are found for the model with no X-rays and low  $\zeta_{H_2}$, the 179.53$\mu m$ line  is reduced by a factor of $\sim 3$ as compared to the fiducial model, while the 538.29$\mu m$ line is only decreased by a factor $\sim 1.5$ as compared to the fiducial model.

\subsection{Cold water emission from parameter survey}

\begin{figure}
    \centering
    \includegraphics[width=0.45\textwidth]{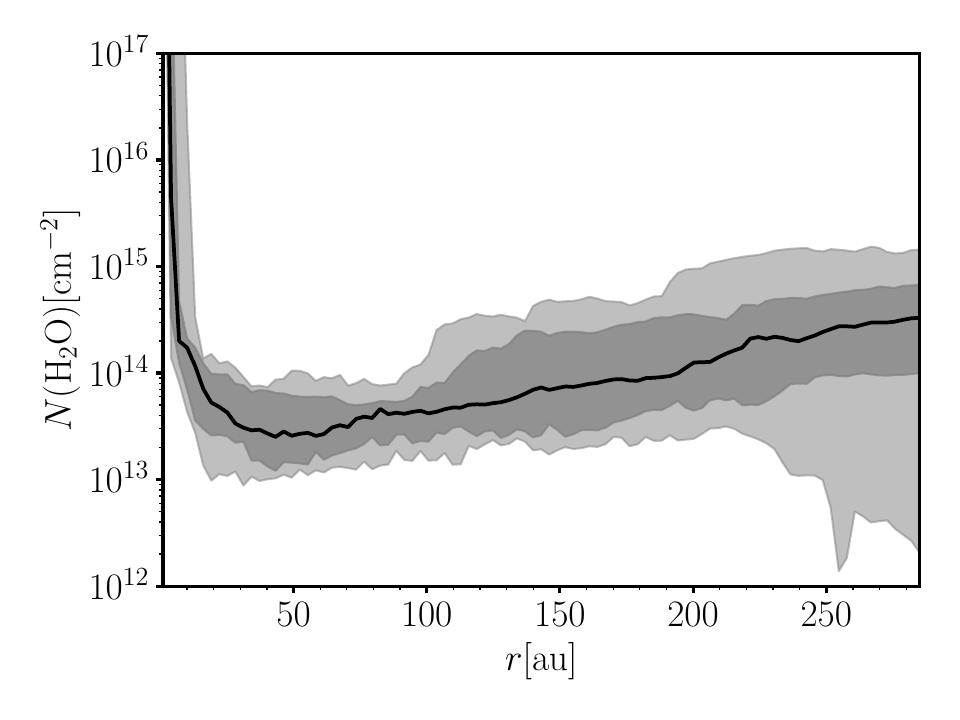}
    \caption{Range of vertical column density of water vapor computed for all the models presented in this work. The black line shows the median of the distribution of the computed vertical column densities. The dark gray filled area shows the 16th and 84th percentiles, while the light gray filled area shows the min and max of the distribution.}
    \label{fig:coldens_h2o}
\end{figure}
\begin{figure*}
    \centering
    \includegraphics[width=\textwidth]{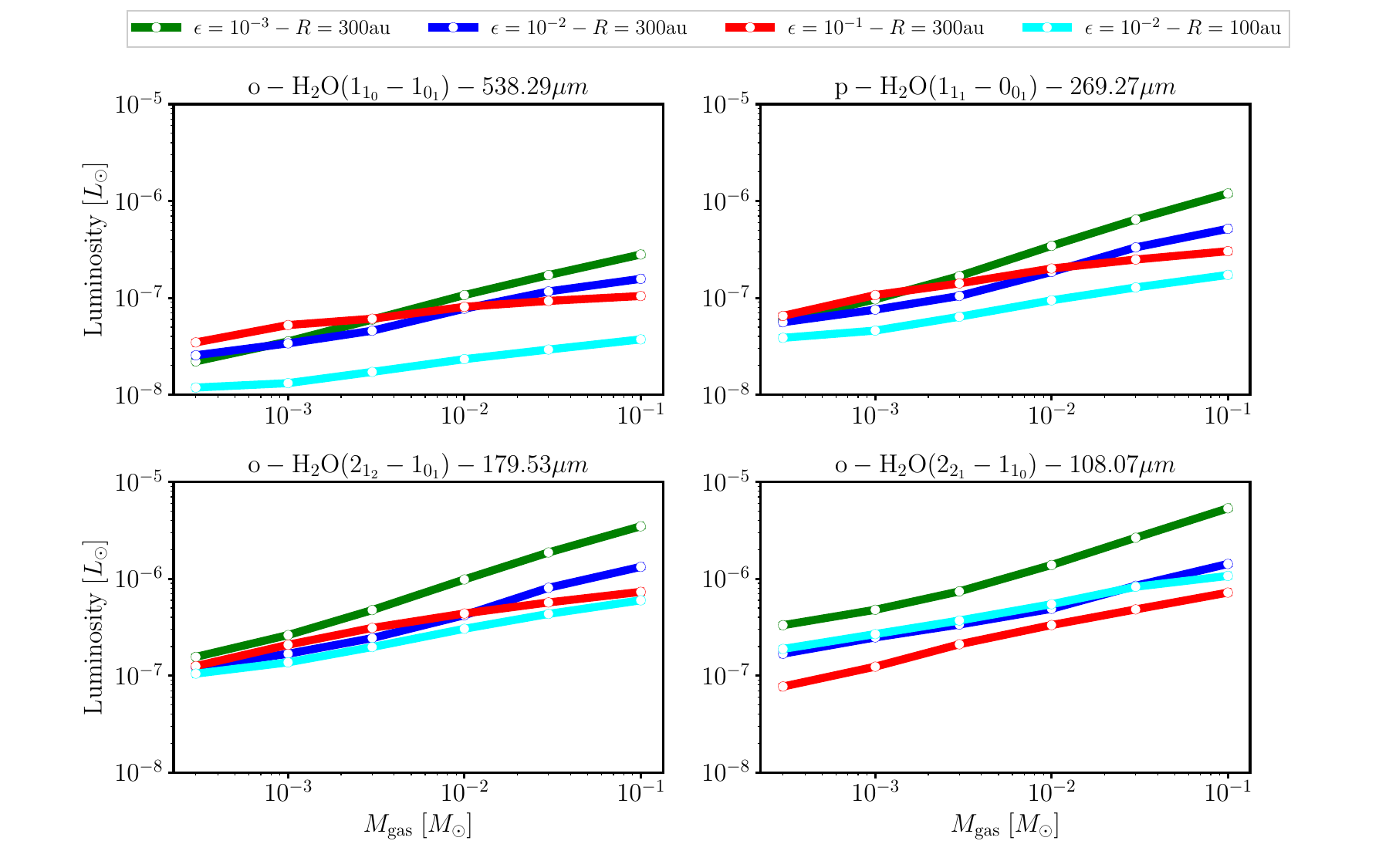}
    \caption{o-H$_2$O ($1_{1_0}-1_{0_1}$), p-H$_2$O ($1_{1_1}-0_{0_0}$), o-H$_2$O ($2_{1_2}-1_{0_1}$) and o-H$_2$O ($2_{2_1}-1_{1_0}$) emission as a function of gas mass for different gas/dust mass ratios ($\epsilon=10^{-3}$: green, $\epsilon=10^{-2}$: blue and $\epsilon=10^{-1}$: red). Cyan color data show the $\epsilon=10^{-2}$ model disks with radii of 100 au and are to be compared with the blue $\epsilon=10^{-2}$ data set with radii of 300 au.}
    \label{fig:water_lum}
\end{figure*}

We now discuss line luminosities computed from all our models, and compare them to \textit{Herschel} observations in the next section. We find that the line luminosities across the parameter survey only vary by approximately an order of magnitude even though the mass range and $\epsilon$  ranges considered span several orders of magnitude. This is due to both the optically thick nature of the transitions and also because the column density of water vapor varies little with mass and dust/gas ratio when all other parameters are held constant.  
Figure \ref{fig:coldens_h2o} shows the range of vertical column density of water vapor computed for all the models (which include variations in $\sigma_H$ via $\epsilon$). For all these models, we used $Y_\mathrm{pd,sH_2O} = 10^{-3}$. 
The computed vertical column densities of water vapor in the outer disk are set by water ice self-shielding, and stay relatively constant, ranging between $N(\mathrm{H_2O})\sim$ few $\times 10^{13}$ and $\sim$ few $\times 10^{14}$ cm$^{-2}$ in regions located outside the radial water snowline.

Figure \ref{fig:water_lum} shows the computed luminosity for H$_2$O for the \textit{Herschel} detected lines at 108.07$\mu m$, 179.53$\mu m$, 269.27$\mu$m and 538.29$\mu m$ for all the models of our limited parameter survey as a function of the gas disk mass. 
For a given model series (same $\epsilon$), increasing the mass of the disk results in higher luminosities, by factors that depend on the excitation temperature of the transition.  In the $3\lesssim A_{\rm V} \lesssim 10$ mag  region where water lines originate, the column density of water vapor remains relatively constant with disk mass as discussed earlier; however the gas temperature in this region differs. Gas and dust are thermally coupled for the more massive disks, while for lower mass disks, gas temperature is lower than that of dust (heating is due to dust collisions, cooling is due to emission by atoms and molecules). The luminosity of the ground state emission of water (i.e. at 538.29$\mu m$ and 269.27$\mu$m) is only weakly dependent on the mass of the disk as well as on the assumed gas/dust mass ratio. The low gas temperatures in the water-emitting region ($T_{gas} \sim 50$ K at 50 au, and $\sim 30$ K at 200 au) make the higher energy emission lines particularly sensitive to the small change in temperature with mass and these lines increase with mass faster than the ground state transitions (see Fig.~\ref{fig:water_lum}). For a given mass, changing $\epsilon$ does not affect the line fluxes very much. In the water emitting region, gas and dust are thermally coupled and the dust temperature only depends on the geometry/disk flaring \citep[see][]{Ruaud19} which is not very sensitive to $\epsilon$.  

\section{Comparison with observations}
\label{sec:comparison}

\begin{figure*}
    \centering
    \includegraphics[width=\textwidth]{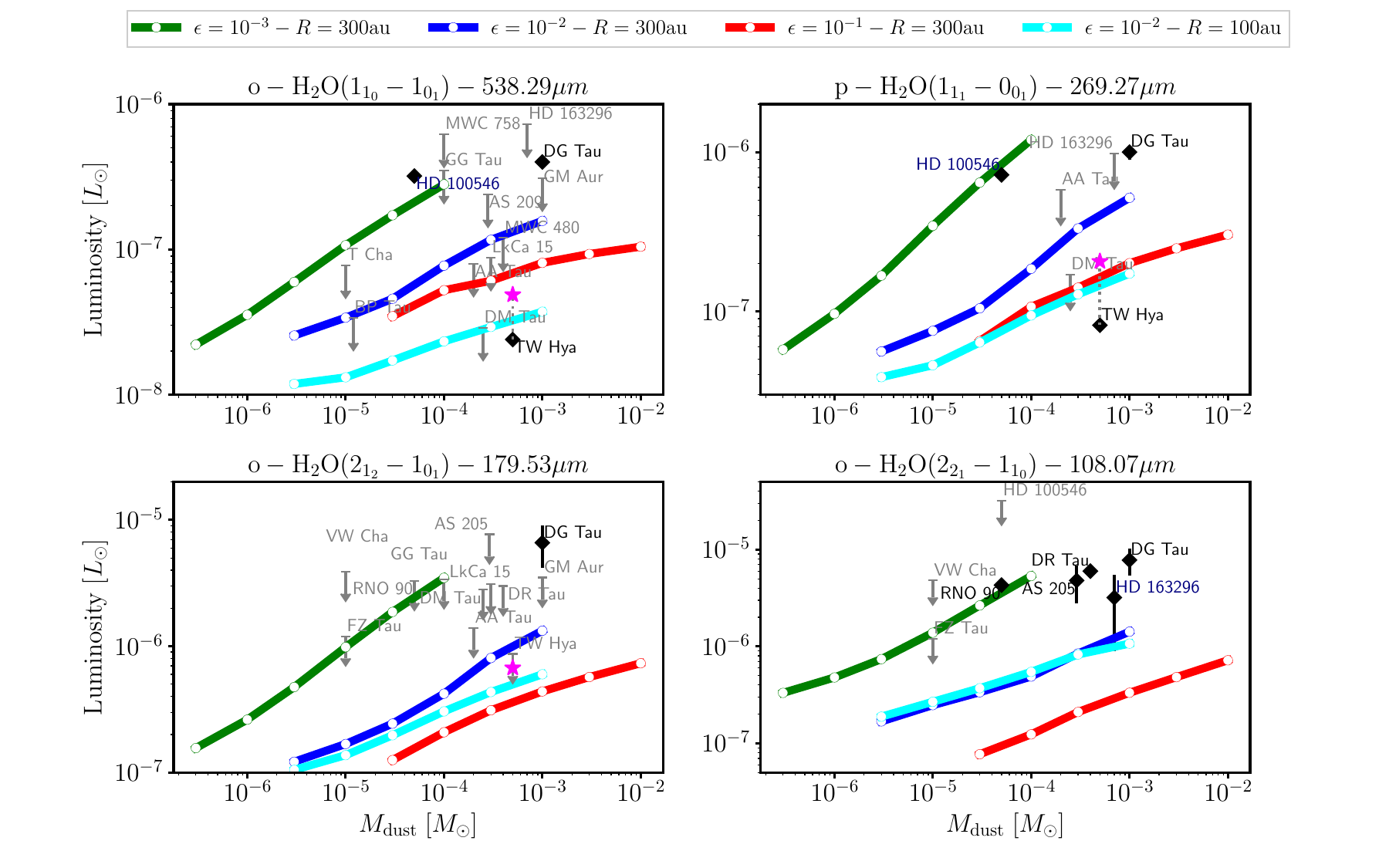}
    \caption{o-H$_2$O ($2_{1_2}-1_{0_1}$), o-H$_2$O ($2_{2_1}-1_{1_0}$), o-H$_2$O ($2_{1_2}-1_{0_1}$) and o-H$_2$O ($2_{2_1}-1_{1_0}$) emission as a function of dust mass for different gas/dust mass ratios and disk radial extent. Data in black show observational constraints obtained by Herschel in different disks. Labels in black indicate disks around T Tauri stars while labels in dark blue indicate disks around Herbig stars. Magenta stars are results obtained with our source specific model of TW Hya  where we used ISM-like oxygen and carbon elemental abundances. Observations in this disk can be explained to within a factor of $\sim 2$.}
    \label{fig:water_lum2}
\end{figure*}

\begin{table*}[]
    \centering
    \begin{tabular}{lccccccc}
    \hline
    \hline
    Source  &    \multicolumn{4}{c}{H$_2$O luminosity ($L_\odot$)} & $M_\mathrm{dust}$ ($M_\odot$) &  $d (pc)$ &  Refs.\\
    \cline{2-5}
            &      538.29$\mu m$    & 269.27$\mu m$ & 179.53$\mu m$ &  108.7$\mu m$ \\
    \hline
TW Hya     & $ 2.4 \pm 0.1 \times 10^{-8}$   &  $ 8.2 \pm 0.5 \times 10^{-8}$ & $<8.7\times 10^{-7}$          &  -                             &  $5.0\times 10^{-4}$  & $ 60$ &  (1),(6)\\      
HD 100546  & $ 3.2 \pm 0.3 \times 10^{-7}$   &  $ 7.2 \pm 0.1 \times 10^{-7}$ &  -                            & $<3.2\times 10^{-5}$           &  $5.0\times 10^{-5}$  & $103$ &   (4)\\            
DG Tau     & $ 4.0 \pm 0.05 \times 10^{-7}$  &  $ 1.0 \pm 0.1 \times 10^{-6}$ & $ 6.6 \pm 2.4 \times 10^{-6}$ & $ 7.8 \pm 2.4 \times 10^{-6}$  &  $1.0\times 10^{-3}$  & $140$ &  (3),(5)\\
AA Tau     & $<8.0\times 10^{-8}$            &  $<5.8\times 10^{-7}$          & $<1.4\times 10^{-6}$          &  -                             &  $2.0\times 10^{-4}$  & $140$ &  (1),(7)\\      
DM Tau     & $<2.9\times 10^{-8}$            &  $<1.7\times 10^{-7}$          & $<2.8\times 10^{-6}$          &                                &  $2.5\times 10^{-4}$  & $140$ &   (1),(7)\\      
HD 163296  & $<7.3\times 10^{-7}$            &  $<9.8\times 10^{-7}$          & -                             & $ 3.2 \pm 2.3 \times 10^{-6}$  &  $7.0\times 10^{-4}$  & $122$ &  (1),(4)\\         
T Cha      & $<7.8\times 10^{-8}$            &                                & -                             & -                              &  $1.0\times 10^{-5}$  & $ 66$ &  (1)\\      
BP Tau     & $<3.4\times 10^{-8}$            &                                & -                             & -                              &  $1.2\times 10^{-5}$  & $ 56$ & (1)\\      
MWC 758    & $<6.2\times 10^{-7}$            &                                & -                             & -                              &  $1.0\times 10^{-4}$  & $200$ &  (1)\\      
GG Tau     & $<3.5\times 10^{-7}$            &                                & $<3.4\times 10^{-6}$          & -                              &  $1.0\times 10^{-4}$  & $140$ &   (1),(7)\\      
AS 209     & $<2.4\times 10^{-7}$            &                                & -                             & -                              &  $2.8\times 10^{-4}$  & $125$ &   (1)\\      
MWC 480    & $<1.2\times 10^{-7}$            &                                & -                             & -                              &  $4.0\times 10^{-4}$  & $131$ &  (1)\\      
LkCa 15    & $<8.8\times 10^{-8}$            &                                & $<3.1\times 10^{-6}$          & -                              &  $3.0\times 10^{-4}$  & $145$ &  (1),(7)\\      
GM Aur     & $<3.1\times 10^{-7}$            &                                & $<3.5\times 10^{-6}$          & -                              &  $1.0\times 10^{-3}$  & $140$ &  (1),(7)\\      
DR Tau     & -                               &  -                             & $<3.0\times 10^{-6}$          & $ 6.0\pm 0.6 \times 10^{-6}$   &  $4.0\times 10^{-4}$  & $140$ &   (2)\\
FZ Tau     & -                               &  -                             & $<1.2\times 10^{-6}$          & $<1.2\times 10^{-6}$           &  $1.0\times 10^{-5}$  & $140$ &   (2)\\      
RNO 90     & -                               &  -                             & $<3.3\times 10^{-6}$          & $ 4.3 \pm 0.5 \times 10^{-6}$  &  $5.0\times 10^{-5}$  & $125$ &    (2)\\   
VW Cha     & -                               &  -                             & $<3.9\times 10^{-6}$          & $<4.8\times 10^{-6}$           &  $1.0\times 10^{-5}$  & $178$ &    (2)\\      
AS 205     & -                               &  -                             & $<7.7\times 10^{-6}$          & $ 4.8 \pm 2.0 \times 10^{-6}$  &  $2.9\times 10^{-4}$  & $125$ &    (3)\\                    
    \hline
    \end{tabular}
    \caption{(1)\citet{Du17}, (2)\citet{Blevins16},(3) \citet{Fedele13}, (4) \citet{Pirovano22}, (5) \citet{Podio13}, (6) \citet{Kamp13}, (7) \citet{Alonso17}.}
    \label{tab:water_obs}
\end{table*}

We next compare our model predictions with observational constraints obtained for  H$_2$O with Herschel-HIFI at $\lambda= 269.27\mu$m and 538.29$\mu m$ and PACS at $\lambda= 108.07 \mu m$ and 179.53$\mu m$ (see Table \ref{tab:water_obs}).

At wavelengths covered by HIFI (i.e. $\lambda=269.27\mu$m and 538.29$\mu m$), only three disks yielded detection; i.e. TW Hya, DG Tau and HD 1005546. For all other disks, only upper limits could be obtained. Using PACS, only DG Tau was detected at  $\lambda=179.53\mu m$ and five disks yielded detection at $\lambda=108.07\mu m$ \citep[i.e. AS 205, DR Tau, DG Tau, RNO 90 and HD 163296][]{Fedele13,Blevins16,Pirovano22}. We note that the sample of objects for which constraints have been obtained for cold water are widely heterogeneous; i.e. covering a wide range of disk surface densities, evolutionary stages, and stellar types. Figure \ref{fig:water_lum} shows a comparison of our model predictions with these \textit{Herschel} observations. Even though the range of parameters explored with our disk parameters survey include changes in the disk masses, disk radial extent and gas/dust mass ratio, the profile of the surface density is fixed. Further, the stellar parameters are fixed (i.e. $M_\mathrm{*} = 1.0 M_\odot$, $R_\mathrm{*} = 2.45 R_\odot$, and UV and X-ray spectra are similar to those of TW Hya). While Fig.~\ref{fig:water_lum2} also shows data from the more massive Herbig disks, these disks are expected to be warmer in general which affects the optically thick water emission. The comparisons between models and observations are therefore more qualitative in nature. 

Figure \ref{fig:water_lum2} demonstrates that all observations including detections and upper limits are consistent with the model range of predictions. We first note that most of the data points lie above the model predictions for models where we have assumed $\epsilon=10^{-2}$, and therefore do not support any depletion of elemental oxygen (discussed in detail in the next section). 

In particular, the observed trends in the data are consistent with the modeling analysis presented thus far; water emission is sensitive to temperature and  \textit{Herschel} constraints support an ISM like oxygen elemental abundance and normal gas/dust mass ratio (i.e., $\epsilon=10^{-2}$).  The three disks with water detections more or less align with model predictions. 
HD 100546 is a Herbig Ae/Be star with a higher UV flux; the gas disk is expected to be at higher temperatures which would drive the emission higher and observed emission is $\sim$ 10 higher than  our models with $\epsilon=10^{-2}$ and $R=300$ au.  DG Tau is a young, accreting star with a strong outflow and  has an order of magnitude higher UV flux \citep{Podio13}  than our TW Hya template; the disk is again thus warmer and our closest models with $\epsilon=10^{-2}$ and $R=100$ au  predict lower fluxes than observations at  $\lambda=108.38\mu m$, $179.38\mu m$ and $269.27\mu m$  and $\lambda=538.29\mu m$ by a factor of $\sim 4-6$.  
For TW Hya, deep observations have been obtained at $\lambda=269.27\mu$m and 538.29$\mu m$. For this disk, we can compare data with the source specific model from \citet{Gorti11} (shown with magenta stars on Fig. \ref{fig:water_lum}). In \citet{Ruaud22}, we show that using this model, observed emission of HD, [O I], [C I] CO isotopologues, C$_2$H and C$_3$H$_2$ can be explained without having to invoke large depletion factors of C and O. The model where we use ISM-like oxygen and carbon elemental abundances can, to within a factor of $\sim$2, explain the emission at $\lambda=538.29\mu m$ and $\lambda=269.27\mu m$.  TW Hya also has rough estimates of the main reservoir in the form of water ice \citep{Debes13}, which corresponds approximately to all gas-phase O not in CO being in  H$_2$O ice, and is consistent with expectations and our models.

There are a few more detections at 108.07$\mu m$, and our models with normal gas/dust ratio underpredict fluxes compared to observed values in all cases. HD 163296 is again a Herbig Ae/Be star, with a warmer disk and expected to be brighter than our model disks. For DR Tau, AS 205 and RNO 90, our models with $\epsilon=10^{-2}$ underpredict observation at 108.07$\mu m$ by a factor of $\sim 10$. DR Tau is a well known variable, high-accretion, T-tauri disk that shows a disk wind signature in CO \citep{Pontoppidan11} and water could therefore be originating from a warmer gas. AS 205 is thought to be relatively young and active disk with indications of a disk wind flow as well \citep{Pontoppidan11,Salyk14}. RNO 90 is thought to be surrounded by a classical disk but it is also known to be have one of the brightest lines in the infrared \citep{Pontoppidan19}, which makes it a peculiar source.

 For disks for which only upper limits were obtained with HIFI and PACS, only DM Tau, for which deep observations have been obtained at 269.27$\mu$m and 538.29$\mu$m, is notably discrepant with the models. While we do not have a model specific to this disk, \citet{Francis22} recently modeled spatially resolved ALMA emission from the DM Tau disk in detail and derive gas and dust masses of $\sim 6\times10^{-3}$ and $\sim 6\times10^{-4}$ M$_{\odot}$ respectively. Their adopted surface density distribution is very similar to our models for the same mass disk, and tapers off at $r=126$au. We can thus compare their best fit masses to the appropriate tracks of our models (red track in Fig.~\ref{fig:water_lum2}).  We then find that the calculated values lie within a factor of two of the upper limits on the water line luminosities. If we additionally consider the fact that DM Tau is less massive and therefore 8 times less luminous than the central star in the models, and also that the disk has a very low PAH abundance according to \citet{Francis22}, we expect the gas temperature to be much lower and result in lower luminosities for the optically thick water lines.

To summarize, our models suggest that the gas temperature is the main parameter affecting the optically thick water emission. Most of the constraints obtained on cold water emission in disks are consistent with an ISM like oxygen elemental abundance (i.e., O/H$=3.2\times 10^{-4}$) and normal gas/dust mass ratio (i.e., $\epsilon=10^{-2}$), with perhaps small variations but  no large scale depletions are needed. We note however that with the sensitivity to temperature and likely dust properties as well,  source-specific models that account disk size, radial drift of dust grains, stellar spectra and masses are likely to bring the agreement with data closer, although this work remains to be addressed in the future.  

\section{Discussion}
\label{sec:discussion}
The results presented so far demonstrate that when thermochemical modeling includes grain surface chemistry, FUV radiative transfer, and self-consistently and iteratively solves for the density structure along with temperature and chemistry, the calculated water emission is consistent with currently available observational data for disks. Our results differ substantially from earlier work in that we do not need to invoke ad hoc depletions of water or O elemental abundance to explain observed water emission. 

There have been limited number of modeling studies targeted at explaining water observations, with the most comprehensive previous studies being those by \citet{Kamp13}, \citet{Du17}, and a more recent study of two Herbig Ae disks using DALI by \citet{Pirovano22}. \citet{Du17} and \citet{Pirovano22} both overproduce water emission. In \citet{Du17} for instance, models using an ISM-like oxygen elemental abundance, a dust-to-gas mass ratio of $10^{-2}$ and a photodesorption yield similar to ours (i.e. $Y_\mathrm{pd,sH_2O} = 10^{-3}$) overpredict the observed emission of water vapor, while the ISM-like parameters could match TW Hya data in \citet{Du14}. \citet{Du17} needed to reduce the water ice abundance in these disk -- and by extension the amount of photodesorbed water -- to match observations. In this paper, they speculate that settling of large icy grains and their migration removes some amount of gas-phase oxygen capable of participating in disk chemistry. Contrary to \citet{Du17}, our models with an ISM-like oxygen elemental abundance and dust-to-gas mass ratio, and $Y_\mathrm{pd,sH_2O} = 10^{-3}$ are consistent with currently available data for disks. We therefore attribute the differences between models to be mainly due to disk thermal structure, although we note that it is difficult to make direct comparisons without a dedicated benchmarking effort.  As we have shown in this paper, far-infrared water emission lines are in general optically thick. The high optical depths lower the effective critical densities, and in addition, densities in the water emitting layers of disks are high (\S \ref{sec:sensitivity}). The optically thick line fluxes are therefore relatively insensitive to the water abundance and more sensitive to gas temperature. We also note that arbitrary depletions in abundances (of elemental O in \citet{Du17} and of water in \citet{Pirovano22}) to reduce the fluxes of optically thick lines will perforce require large depletion factors. To reconcile with observed data, \citet{Du17} needed to decrease the initial abundance of oxygen by a factor of 100 to $10^4$ (with $\epsilon=10^{-2}$). \citet{Pirovano22} needed to parameterize water abundance to match emission data, and these water abundances are roughly a factor of $\sim 10$ lower than our calculated values for T Tauri disks. 

Identifying reasons for the possible difference on the computed disk thermal structure is difficult because many processes are involved in setting the disk thermal balance (i.e. cooling for instance is by lines and therefore depends on chemistry). Also, unlike other models, we do iterate between the density, chemistry and temperature at each spatial position to maintain vertical hydrostatic equilibrium. This will affect photon penetration to the desorption layer and affect the disk physical structure among other things, as also noted in \citet{Du14}.  Important differences with \citet{Du17} include our considerations of FUV shielding by gas and ice species, in particular water ice, the details of the dust disk structure and resulting cross-sectional area, and our more detailed treatment of cosmic ray and X-ray ionization processes. These are all likely to result in subtle differences in the thermal structure affecting the computed line luminosities; as we have argued thus far, the column density of water itself does not significantly vary with the various model parameters due to water self-shielding and neither does the optically thick emission itself depend significantly on the column density of water.  

Since our model parameter survey is for disks around solar-mass stars, our results cannot be directly compared with the Herbig Ae disk modeling of \cite{Pirovano22}. We expect the column density of desorbed water to be similar, due to the weak dependence on UV flux (see Eq.~\ref{eq:h2o_sat}), while gas temperatures are expected to be higher due to increased heating around more massive stars, which would affect the line luminosities. 

Our results are more in accord with the results presented in \citet{Kamp13} where they study the impact of various disk parameters and processes, such as X-ray chemistry, grain surface chemistry and photodesorption on the formation of water in the disk around TW Hya. Similarly to our models, they find that cold water emission is optically thick and that it can be explained with carbon and oxygen elemental abundances typical from the ISM in this disk. They also show that water emission is only weakly dependent onto several disk parameters which we attribute to the optically thick nature of the emission. Therefore, ProDiMo models are consistent with our findings. Finally, our conclusions are also in agreement with the conclusions made by  \citet{Podio13} when modeling Herschel emission from DG Tau using ProDiMo; they were able to reproduce water emission to within a factor of two without resorting to depletion of oxygen or water, and also concluded that the water emission lines were optically thick due to the high densities in the emitting layers. 

\section{Implications}
Depletion of water is often attributed to underlying planet formation processes at work in disks \citep[e.g.,][]{Krijt16, Du17, Pirovano22}. Here planetesimal formation locks water ices into larger bodies and essentially removes a reservoir of oxygen (or water) from participating in disk chemistry; the gas phase chemical species subsequently diffuse across the chemical gradient from surface layers to regions below the vertical snowlines and get sequestered into ices that get incorporated into planetesimals.  Over time this is expected to result in depletion factors of several orders of magnitude of various species involving C and O \citep{Bergin16, Krijt18, Oberg21}. 

Here, we contend that observed line emission of many species including CO \citep{Ruaud22}, [CI] \citep{Pascucci2023} and water (this work) do not support any large scale depletions of elemental species as has been previously claimed. Why does the diffusion of gas phase species not lead to depletion? Since the ground state water and most sub-millimeter emission lines originate from cold regions of the disk close to the vertical water snowline \citep[][this work]{Ruaud19}, and since grain surface chemistry plays an important role in dictating the strength of the observed emission, dynamical models of disks need to account for gas-grain interactions and dust particle dynamics in a consistent manner with the chemical evolution. Such comprehensive models are not yet available. Preliminary results from a full 2D chemodynamical simulation including grain growth and dynamics (Ruaud \& Gorti, in preparation) indicate that inclusion of the vertical diffusion of small grains returns sequestered ices in the midplane regions back to the surface. Therefore, when diffusion of gas and dust grains are both considered, there is no large scale depletion of chemical species from the surface layers as found earlier. 

Related to the above issue is the role of grain surface chemistry on the sequestration of carbon and oxygen from the gas and the setting of a vertical C/O gradient. In \citet{Ruaud19}, we found that the main gas phase oxygen and carbon reservoirs, namely atomic oxygen and CO, condensed near the vertical water snowline and that CO freezes out as CO$_2$ on grains in the exposed layer of the disk midplane. This is due to an efficient conversion of CO into CO$_2$ ice by photoprocessing of water ice. This efficient conversion was also found in other models, 
e.g. \citet{Aikawa15,Eistrup16,2018A&A...618A.182B}. In our model, this implies that freeze-out in the outer disk does not result in a change in the C/O ratio (the elemental abundance of oxygen is approximately twice that of carbon). While preliminary results from JWST support this picture of mixed ices made of water, CO and CO$_2$ \citep{sturm2023}, this is still an open question. There are still many unresolved issues such as the porosity of dust grains which could affect the penetration of UV photons (and affect the conversion of CO to CO$_2$ ice) into the ice layers, and the many uncertain  parameters (e.g., binding and diffusion energies of many key species) that enter in the chemical modeling that could impact the efficiency of the resulting chemistry.

We note that the present modeling results do not rule out smaller changes (within a factor of a few) in the bulk elemental abundances in the surface layers or in their ratios. Specifically for water, which is the focus of this paper, we argue that the lines are mostly optically thick and cannot be used to infer changes in bulk elemental ratios based on weak emission.

\section{Conclusions} 
\label{sec:conclusions}

Using detailed equilibrium disk thermo-chemical models coupled with full three phase gas-grain chemical model we investigate cold water emission observed with Herschel. We use models from a previous disk parameter survey where we study the effect of varying gas disk masses, gas/dust mass ratio, and disk outer radius \citep{Ruaud22}. We focus on the cold water lines originating from the outer disk (i.e. with upper energy levels $E_\mathrm{up}\lesssim200$ K). Our conclusions are as follows:
\begin{itemize}
    \item Cold water emission originates from a  narrow region located at the interface between the molecular layer and the disk midplane and extends uniformly throughout the disk. This water is mainly produced by X-ray induced ion-chemistry in the inner disk and photodesorption of water ice in the outer disk.
    \item In regions where photodesorption dominates, because the formation and destruction rates of water vapor both depend linearly on the local UV flux, the abundance of desorbed water vapor does not depend on the UV flux but saturates at $\sim 10^{-8}$, for a photodesorption yield of $10^{-3}$ molecules per photons and a dust cross sectional area per H atom of $\sigma_\mathrm{H} = 10^{-22}$ cm$^2$. The dust cross-sectional area per H atom and the photodesorption yield both affect this saturation abundance. 
    \item The cold water emission computed in all our models is mostly optically thick and therefore mainly depends on the computed disk thermal structure.
    \item The total vertical column density of cold water vapor is most affected by X-ray ionization, the photodesorption yield and chemical desorption. Water vapor production is insensitive to the cosmic ray ionization rate. Although varying parameters such as the photodesorption yield or taking into acccount processes like chemical desorption change the computed vertical column of water vapor it does not impact the predicted line luminosities. This is due to the optically thick nature of the emission.
    \item Cold water emission is weakly dependent on disk parameters such as gas mass and gas/dust ratio. The computed vertical column densities of water varies over a limited range in between $N(\mathrm{H_2O})\sim$ few $\times 10^{13}$ and $\sim$ few $\times 10^{14}$ cm$^{-2}$ (outside the radial water snowline) even when the gas mass and gas/dust ratio are varied by 3 orders of magnitude.
    \item Our modeling results do not indicate large scale depletion factors of oxygen or water in disks. A comparison of our modeling results with constraints obtained with Herschel HIFI and PACS show that most of the observations are consistent with disks having ISM like oxygen elemental abundance and gas/dust mass ratios.  
    \item While our results are in agreement with previous work using the ProDiMo model \citep{Podio13, Kamp13} where the water lines were found to be optically thick, we do not concur with other modeling efforts that call for large scale depletion of oxygen or water \citep{Du17, Pirovano22}.  We conclude that due to the optically thick nature of cold water emission, the line fluxes cannot be used to infer or constrain elemental O/H ratios in disks.
\end{itemize}

We are thankful to the anonymous referee who provided comments that improved this contribution. MR acknowledges support from NASA/EW research grant 80NSSC21K0391 which made this work possible. Support for UG's research was also provided by NASA’s Planetary Science Division Research Program, through ISFM work package ‘The Production of Astrobiologically Important Organics during Early Planetary System Formation and Evolution’ at NASA Ames Research Center.

\bibliography{bibliography}
\bibliographystyle{aasjournal}

\end{document}